# First Order Approximations of the Pythagorean Won-Loss Formula for Predicting MLB Teams' Winning Percentages

May 21, 2012
By Kevin D. Dayaratna and Steven J. Miller[*]

## Abstract

We mathematically prove that an existing linear predictor of baseball teams' winning percentages (Jones and Tappin 2005) is simply just a first-order approximation to Bill James' Pythagorean Won-Loss formula and can thus be written in terms of the formula's well-known exponent. We estimate the linear model on twenty seasons of Major League Baseball data and are able to verify that the resulting coefficient estimate, with 95% confidence, is virtually identical to the empirically accepted value of 1.82. Our work thus helps explain why this simple and elegant model is such a strong linear predictor.

## I. Introduction

First postulated by Bill James in the early 1980s, the Pythagorean Won-Loss formula indicates the percentage of games (winning percentage, WP) a baseball team should have won at a particular point in a season as a function of average runs scored (RS) and average runs allowed (RA):

$$\text{WP} \approx \frac{\text{RS}^\gamma}{\text{RS}^\gamma + \text{RA}^\gamma}.$$

James initially postulated the exponent $\gamma$ to be 2 (hence the name "Pythagorean" from a sum of squares). Empirical observation suggested that $\gamma \approx 1.8$ was more appropriate.

For decades, the Pythagorean Won-Loss formula gave a strong indication of the percentage of games a baseball team should have won at a particular point in a season. Until just a few years ago, however, the formula had no statistical verification. Miller (2007) provided such verification by assuming that runs scored and runs allowed follow separate independent continuous Weibull distributions. Upon making these assumptions, he was able to derive James's formula in the form of the probability that the runs a particular team scores is greater than the runs it allows. He estimated this model via least squares and maximum likelihood estimation on 2004 American League data and determined that the appropriate value of $\gamma$ was indeed around 1.8, consistent with empirical observation.

Jones and Tappin (2005) presented a simple linear model that also serves as a predictor of the team's winning percentage. In the following section, we prove that this formula is actually nothing but a first order approximation to the Pythagorean Won-Loss formula:

---

[*] Kevin D. Dayaratna (kevind@math.umd.edu) is a graduate student in Mathematical Statistics at the University of Maryland. Steven J. Miller (steven.j.miller@williams.edu) is an Associate Professor of Mathematics and Statistics at Williams College. The second author is partially supported by NSF Grant DMS0970067. We would like to thank Eric Fritz for writing a script to download baseball data from BaseballAlmanac.com, Armaan Ambati for help with the data, and Ben Baumer for helpful comments on an earlier draft.

$$WP = .500 + \beta(RS - RA);$$

here $WP$ is the team's winning percentage, $RS$ is the average points scored (goals in hockey, runs in baseball, et cetera) and $RA$ is the average points allowed. Notice that if $RS = RA$ then the team is predicted to win half its games. Typically $\beta$ is a small number. As a result, for observed values of $RS$ and $RA$ we do not need to worry about the above expression exceeding 1.000 or falling below 0.000. For example, in baseball in 2010 runs scored ranged from 513 to 859 and runs allowed from 581 to 845. For these ranges, the winning percentages are all `reasonable,' ranging from 0.352 to 0.599. (MLB.com)

## II. Derivation of Linear Predictor

We now show how the above linear predictor follows from the Pythagorean formula. We assume there is some exponent $\gamma$ such that

$$WP = \frac{RS^\gamma}{RS^\gamma + RA^\gamma}.$$

We provide a simple statistical derivation of the linear formula utilizing multivariable calculus.

Proof:

In this subsection we assume the reader is familiar with multivariable calculus. Recall the second order Taylor series expansion of a function $f(x, y)$ about the point $(a,b)$ is

$$f(x, y) = f(a,b) + \frac{\partial f}{\partial x}\bigg|_{(a,b)} (x-a) + \frac{\partial f}{\partial y}\bigg|_{(a,b)} (y-b) + \frac{1}{2}\frac{\partial^2 f}{\partial x^2}\bigg|_{(a,b)} (x-a)^2$$
$$+ \frac{\partial^2 f}{\partial x \partial y}\bigg|_{(a,b)} (x-a)(y-b) + \frac{1}{2}\frac{\partial^2 f}{\partial y^2}\bigg|_{(a,b)} (y-b)^2$$
$$+ \text{higher order terms}.$$

Here, the higher order terms involve products of $(x-a)$ and $(y-b)$ to the third and higher powers. The tangent plane approximation, which means keeping just the constant and linear terms, is

$$f(x, y) = f(a,b) + \frac{\partial f}{\partial x}\bigg|_{(a,b)} (x-a) + \frac{\partial f}{\partial y}\bigg|_{(a,b)} (y-b).$$

Let $R_{ave}$ denote the average number of runs scored in the league. We let

$$f(x, y) = \frac{x^\gamma}{x^\gamma + y^\gamma}.$$

We now expand about the point $(a,b) = (R_{ave}, R_{ave})$, with $x = RS$ and $y = RA$, so

$$f(R_{ave}, R_{ave}) = .500$$

$$\frac{\partial f}{\partial x} = \frac{\gamma x^{\gamma-1} y^{\gamma}}{(x^{\gamma} + y^{\gamma})^2} \quad \Rightarrow \quad \left.\frac{\partial f}{\partial x}\right|_{(R_{ave}, R_{ave})} = \frac{\gamma}{4R_{ave}}$$

$$\frac{\partial f}{\partial y} = -\frac{\gamma x^{\gamma} y^{\gamma-1}}{(x^{\gamma} + y^{\gamma})^2} \quad \Rightarrow \quad \left.\frac{\partial f}{\partial y}\right|_{(R_{ave}, R_{ave})} = -\frac{\gamma}{4R_{ave}}.$$

Noting that the predicted winning percentage is $f(RS, RA)$, we see that the first order, multivariate Taylor series expansion about $(RS, RA)$ gives

$$WP \approx .500 + \frac{\gamma}{4R_{ave}}(RS - R_{ave}) - \frac{\gamma}{4R_{ave}}(RA - R_{ave}) = .500 + \frac{\gamma}{4R_{ave}}(RS - RA).$$

### III. Model Estimation

Michael Jones and Linda Tappin (2005) used this linear model for baseball. They wrote $WP = .500 + \beta(RS - RA)$, and by looking at the seasonal data from 1969 to 2003 found the best values of $\beta$ ranged from .00053 to .00078, with an average value of .00065. Taking their average value of .00065 and using $\gamma = 1.81$ leads to a predicted value of 696 runs scored per team per year, or about 4.3 runs per game. Conversely, using the average number of runs scored in 2010 by American League teams (721) and their average value of $\beta$, one gets a prediction of 1.88 for $\gamma$.

Our analysis in Section II provides theoretical support for the linear model. In particular, the slope is no longer a mysterious quantity, but is naturally related to the exponent and average scoring in the league. Here, we also provide empirical support by estimating the model via the method of least squares:

$$WP \approx \alpha + \beta(RS - RA), \text{ where } \beta = \frac{\gamma}{4R_{ave}}.$$

Below are our estimates via the method of least squares:

Coefficient Estimates and Model Fit Statistics

| Season | $\hat{\alpha}$ | $\hat{\beta}$ | $R_{ave}$ | $\hat{\gamma}$ | 95% Lower Bound on $\hat{\gamma}$ | 95% Upper Bound on $\hat{\gamma}$ | $R^2$ |
|---|---|---|---|---|---|---|---|
| 1991 | 0.500 | 0.119 | 4.308 | 2.058 | 1.807 | 2.310 | 0.922 |
| 1992 | 0.500 | 0.126 | 4.117 | 2.076 | 1.710 | 2.442 | 0.851 |
| 1993 | 0.500 | 0.109 | 4.598 | 2.001 | 1.645 | 2.359 | 0.851 |
| 1994 | 0.500 | 0.084 | 4.923 | 1.658 | 1.366 | 1.951 | 0.836 |
| 1995 | 0.500 | 0.094 | 4.847 | 1.826 | 1.466 | 2.185 | 0.807 |
| 1996 | 0.500 | 0.091 | 5.036 | 1.825 | 1.564 | 2.085 | 0.889 |

| 1997 | 0.500 | 0.087 | 4.767 | 1.668 | 1.345 | 1.991 | 0.813 |
| 1998 | 0.500 | 0.098 | 4.790 | 1.881 | 1.667 | 2.095 | 0.920 |
| 1999 | 0.500 | 0.099 | 5.085 | 2.010 | 1.794 | 2.226 | 0.929 |
| 2000 | 0.500 | 0.092 | 5.140 | 1.893 | 1.626 | 2.160 | 0.883 |
| 2001 | 0.500 | 0.104 | 4.775 | 1.978 | 1.743 | 2.215 | 0.913 |
| 2002 | 0.500 | 0.103 | 4.618 | 1.908 | 1.682 | 2.134 | 0.914 |
| 2003 | 0.500 | 0.103 | 4.728 | 1.949 | 1.716 | 2.181 | 0.913 |
| 2004 | 0.500 | 0.109 | 4.814 | 2.108 | 1.843 | 2.374 | 0.905 |
| 2005 | 0.500 | 0.095 | 4.586 | 1.737 | 1.436 | 2.040 | 0.833 |
| 2006 | 0.500 | 0.098 | 4.858 | 1.901 | 1.567 | 2.235 | 0.829 |
| 2007 | 0.500 | 0.085 | 4.797 | 1.640 | 1.330 | 1.951 | 0.807 |
| 2008 | 0.500 | 0.104 | 4.651 | 1.931 | 1.619 | 2.244 | 0.851 |
| 2009 | 0.500 | 0.106 | 4.613 | 1.963 | 1.642 | 2.284 | 0.848 |
| 2010 | 0.500 | 0.094 | 4.366 | 1.634 | 1.489 | 1.780 | 0.950 |
| 2011 | 0.500 | 0.104 | 4.283 | 1.775 | 1.506 | 2.045 | 0.867 |

After choosing a standard significance level of 0.05 and instituting Bonferroni corrections, which reduces our significance level to 0.0025, each of our coefficient estimates, as well as overall model fit, are highly significant. This statistical significance, coupled with our coefficients of determination being reasonably close to one, signify that our linear model fits quite well.

Furthermore, with the exception of the 2010 season, the commonly accepted value of 1.82 for the exponent for the Pythagorean Won Loss formula (Miller 2007), falls within all of our 95% confidence intervals. Bonferroni corrections increase the size of all of our confidence intervals, including for the estimates pertaining to the 2010 season (to an interval of [1.399, 1.870]). These facts provide us with empirical verification that the Jones and Tappin (2005) linear model of winning percentages is simply just a first order approximation to the Pythagorean Won-Loss formula.

## IV. Conclusions and Future Research

We have provided a theoretical justification for an existing linear model that allows for an interpretation of the slope parameter in terms of the Pythagorean Won-Loss formula's coefficient. Our theoretical work, along with our model estimation, helps explain why this simple and elegant linear model is such a strong linear predictor.

There are a number of potential avenues of future research we hope our work will encourage. We have presented a first order approximation of the Pythagorean Won-Loss Formula. In future research, one could compare higher order approximations to the one presented here. Secondly, one could examine slight variations in $\gamma$ as a result of changes over time such as steroid use, height of the pitcher's mound, players' diets, and the introduction of inter-league play among others. Thirdly, one could apply this model to other sports such as basketball, hockey, football, and soccer. Finally, it could be fascinating to apply this model to a much larger span of data and compare resulting coefficient estimates for teams of different eras.

## VI. Appendix

Alternative Proof (using single variable calculus):

Recall that we assume there is some exponent $\gamma$ such that

$$WP = \frac{RS^\gamma}{RS^\gamma + RA^\gamma}.$$

We multiply the right hand side by $(1/RS^\gamma)/(1/RS^\gamma)$ and find

$$WP = \frac{1}{1+\frac{RA^\gamma}{RS^\gamma}} = \left(1+\left(\frac{RA}{RS}\right)^\gamma\right)^{-1}.$$

There are many ways to attack the algebra. In the analysis below, we consistently replace complicated functions by their linear approximations (i.e., their first order Taylor series). Inspired by the logit model, let $u_{RS} = \ln(RS)$ and $u_{RA} = \ln(RA)$, so $RS = \exp(u_{RS})$ and $RA = \exp(u_{RA})$. Then $(RA/RS)^\gamma = (\exp(u_{RA})/\exp(u_{RS}))^\gamma$, which is $\exp(-\gamma(u_{RS} - u_{RA}))$. We thus have

$$WP = (1 + \exp(-\gamma(u_{RS} - u_{RA})))^{-1}.$$

We now make some approximations. While there will obviously be some loss in predictive power from these choices, it will lead to a very simple, final expression. As we expect $RS$ and $RA$ to be of comparable size, the difference of their logarithms ($u_{RS} - u_{RA}$) should be small; for example, if $RS = 800$ and $RA = 600$ (reasonable numbers in baseball), one finds $u_{RS} - u_{RA} \approx .288$. We Taylor expand the exponential function, noting

$$\exp(x) = 1 + x + \text{higher order terms.}$$

We drop these higher order terms and we have $x = -\gamma(u_{RS} - u_{RA})$. In other words, we are only keeping the constant and linear terms; note that if we only kept the constant term, there would be no dependence on points scored or allowed! We thus find

$$WP \approx (1 + 1 - \gamma(u_{RS} - u_{RA}))^{-1} = \frac{1}{2}\left(1 - \frac{\gamma}{2}(u_{RS} - u_{RA})\right)^{-1}.$$

We now expand using the geometric series formula, which says

$$\frac{1}{1-r} = 1 + r + r^2 + r^3 + \cdots$$

for $|r| < 1$. We take $r = \frac{\gamma}{2}(u_{RS} - u_{RA})$, and again only keep the constant and linear term, $1+r$, yielding

$$WP \approx \frac{1}{2}\left(1 + \frac{\gamma}{2}(u_{RS} - u_{RA})\right) = \frac{1}{2} + \frac{\gamma}{4}(u_{RS} - u_{RA}).$$

We need to do a little more analysis to obtain a formula that is linear in $RS - RA$. Recalling that the u's are the logarithms of the points, we have

$$u_{RS} - u_{RA} = \ln(RS) - \ln(RA) = \ln\left(\frac{RS}{RA}\right) = \ln\left(\frac{RA + RS - RA}{RA}\right) = \ln\left(1 + \frac{RS - RA}{RA}\right).$$

We now Taylor expand the logarithm. We have $\log(1 + x) = x$ plus higher order terms. For us, $x = \frac{RS - RA}{RA}$ is much less than 1, and thus we again only keep up to the linear term. Substituting yields

$$WP \approx \frac{1}{2} + \frac{\gamma}{4}\frac{RS - RA}{RA}.$$

We make one last simplification. To first order, the $RA$ in the denominator can be replaced by $R_{ave}$, the average number of points scored in the league. We have (finally) reached our linear approximation,

$$WP \approx 0.500 + \frac{\gamma}{4R_{ave}}(RS - RA).$$

Thus, in the simple linear approximation model, the `interesting' coefficient should be approximately $\frac{\gamma}{4R_{ave}}$.